\documentclass{article}

\usepackage{PRIMEarxiv}

\usepackage[utf8]{inputenc} 
\usepackage[T1]{fontenc}    
\usepackage{hyperref}       
\usepackage{url}            
\usepackage{booktabs}       
\usepackage{amsfonts}       
\usepackage{nicefrac}       
\usepackage{microtype}      
\usepackage{lipsum}
\usepackage{fancyhdr}       
\usepackage{graphicx}       
\graphicspath{{media/}}     
\usepackage{abstract}

\usepackage{amsmath}
\usepackage{algorithm}
\usepackage{algpseudocode}
\usepackage{bm}
\algblockdefx[Foreach]{Foreach}{EndForeach}[1]{\textbf{foreach} #1 \textbf{do}}{\textbf{end foreach}}%
\algblockdefx[DoWhile]{Do}{EndDoWhile}{\algorithmicdo}[1]{\algorithmicwhile\ #1}%

\title{MPVNN: Mutated Pathway Visible Neural Network architecture for interpretable prediction of cancer-specific survival risk
}

\author{
  Gourab Ghosh Roy \\
  School of Computer Science \\ 
  University of Birmingham \\
  Birmingham, B15 2TT, UK \\
  School of Computing and Information Systems \\ University of Melbourne \\
  Melbourne, 3052, Australia \\
  \texttt{g.ghoshroy@pgr.bham.ac.uk} \\
   \And
  Nicholas Geard \\
  School of Computing and Information Systems \\ University of Melbourne \\
  Melbourne, 3052, Australia \\
  \texttt{nicholas.geard@unimelb.edu.au } \\
  \And
  Karin Verspoor \\
  School of Computing Technologies \\
  RMIT University \\
  Melbourne, 3000, Australia \\
  \texttt{karin.verspoor@rmit.edu.au } \\
  \And
  Shan He \\
  School of Computer Science \\ 
  University of Birmingham \\
  Birmingham, B15 2TT, UK \\
  \texttt{s.he@cs.bham.ac.uk} \\
}

\begin{document}
\maketitle

\begin{abstract}
Survival risk prediction using gene expression data is important in making treatment decisions in cancer. Standard neural network (NN) survival analysis models are black boxes with lack of interpretability. More interpretable visible neural network (VNN) architectures are designed using biological pathway knowledge. But they do not model how pathway structures can change for particular cancer types. We propose a novel Mutated Pathway VNN or MPVNN architecture, designed using prior signaling pathway knowledge and gene mutation data-based edge randomization simulating signal flow disruption. As a case study, we use the PI3K-Akt pathway and demonstrate overall improved cancer-specific survival risk prediction results of MPVNN over standard non-NN and other similar sized NN survival analysis methods. We show that trained MPVNN architecture interpretation, which points to smaller sets of genes connected by signal flow within the PI3K-Akt pathway that are important in risk prediction for particular cancer types, is reliable.
\end{abstract}


\section{Introduction}

Cancer is a leading cause of death worldwide, and a substantial amount of medical research is focused on survival analysis of cancer patients. The suitability of a particular treatment method could be guided by the predicted survival risk of the patient. The effectiveness of a treatment method could also be measured using the predicted risk. Gene expression data has been extensively used for cancer risk prediction, and different machine learning methods have been applied for this survival analysis task by learning from survival data \cite{cheng2013development}. Neural networks (NNs) are an effective category of machine learning methods which have been used for gene expression-based cancer risk prediction \cite{katzman2018deepsurv,huang2019salmon}.

One major challenge of some machine learning models including standard NNs is lack of interpretability. Model interpretability refers to the degree to which the model's internal operations can be understood by a human \cite{biran2017explanation}, which here denotes being understood in biological terms. This understanding would point to how particular biological entities and relationships are used internally by the model in mapping the input to the output. A more interpretable model would inherently increase a user's trust on the model, and hence be considered more reliable for use in high-stakes clinical applications \cite{rudin2019stop}.

Visible NNs or VNNs are neural networks where biological meanings are attached to intermediate neurons, and these VNNs provide increased model interpretability compared to standard black box NNs \cite{michael2018visible}. Under a visible architecture, neurons represent biological entities like genes, proteins, pathways, cell subsystems, etc. and connections between the neurons represent biological relationships. Here we refer to those NNs as VNNs where neurons in each intermediate layer are associated with some explicit biological meaning.

An important source of biological
knowledge for use in the design of VNNs is that of biological pathways. VNNs whose design uses pathway knowledge with neurons representing pathways include P-NET \cite{elmarakeby2021biologically}, GenNet \cite{van2020gennet} and DCell \cite{ma2018using}. These VNNs are designed from hierarchical knowledge and have multiple intermediate layers. Another VNN designed using pathway knowledge is knowledge-primed NN or KPNN \cite{fortelny2020knowledge}, which is a NN architecture analogous to the structure of signaling pathways and also has multiple intermediate layers. A VNN with one intermediate layer, designed using protein–protein (PPI) and protein–DNA (PDI) interactions data, for predicting cell type from single cell expression values has been explored \cite{lin2017using}. However, none of these VNNs model how a known biological pathway structure can change for a particular disease.

In this study, we propose a Mutated Pathway Visible Neural Network or MPVNN architecture for cancer-specific survival risk prediction from gene expression data. This is a simple VNN architecture, with one intermediate layer between the input genotype and the output phenotype layers. Each neuron in the single intermediate layer is assigned to represent perturbation at a gene. Here a gene is considered to be perturbed if it is in the path of signal flow. A gene perturbation is derived from the expression of the same gene and the expression of the genes it shares edges with. Initially these edges are signaling pathway edges from prior knowledge. However, the signal flow through the known pathway edges can be disrupted in cancer. Simulating the change in pathway structure for a particular cancer type, we use additional gene mutation data for that cancer to replace a certain fraction of known pathway edges with random gene connections. These new connections are aimed at capturing signaling interactions not present in the used prior knowledge, which can be important in survival risk prediction for that particular cancer type. So the connections between the input layer and the intermediate layer neurons are used to represent the signaling edges, either obtained from prior knowledge or from the randomization using mutation data. Finally, from these gene perturbations, the cancer-specific survival risk is predicted in the MPVNN output.

Using prior knowledge of PI3K-Akt signaling pathway and mutation data in the design, we assess the cancer-specific risk prediction performance of our proposed MPVNN architecture on data for 10 cancer types from The Cancer Genome Atlas or TCGA \cite{tomczak2015cancer}. We compare its prediction performance against one standard non-NN and other similar sized NN survival analysis methods. We show how interpretation of trained MPVNN architecture points to signal flow involving gene sets within the larger PI3K-Akt pathway which are important in cancer-specific risk prediction. We also investigate how the interpreted signal flow can be through some connections not present in the used PI3K-Akt pathway prior knowledge. To assess the reliability of the MPVNN interpreted insights for two cancer types, ovarian and liver cancer, we use validation from prior studies in literature. All of this is aimed at evaluating the suitability of MPVNN for use as a cancer-specific survival risk predictor in clinical applications.

Our main contributions in this paper are summarized as follows:
\begin{enumerate}
    \item A novel Mutated Pathway Visible Neural Network or MPVNN architecture designed using signaling pathway knowledge and mutation data-based edge randomization is proposed for cancer-specific survival risk prediction. 
    \item Using the PI3K-Akt signaling pathway as a case study, we demonstrate the overall improved risk prediction performance on real-world cancer datasets for MPVNN over standard and other comparable NN survival analysis methods. 
    \item We also show how MPVNN interpretation can provide insights on signal flow important in risk prediction for particular cancer types, and assess the reliability of some such insights using evidence from literature. 
\end{enumerate}


\section{Methods}

\subsection{Problem}

We address the survival analysis task as a regression problem of predicting the survival risk. There are other approaches of addressing survival analysis. For example, in a classification setting, the patients can be stratified into risk categories \cite{chen2014risk}. However, such a stratification, based on a threshold like the median survival time, might not be accurate in a patient population with a mix of cancer stages corresponding to widely varying medians. In another analysis approach, time-to-event distributions can be estimated \cite{chapfuwa2018adversarial}. Here our aim is to predict risk scores indicative of the survival times so that ordering is correct. A clinician might be interested only in knowing if a particular treatment causes an increase in survival time without the need to know the exact value of the said time. 

The survival data comprises two major components : the $i^{th}$ patient survival time to event of interest $t_i$, and the censoring indicator $l_i$ where a value of $1$ denotes that the event was observed and a value of $0$ denotes censoring. In our risk prediction task, the objective is to predict a relative risk score $f(x_i)$ as output given the expression profile $x_i \in \mathbb{R}^{N}$ consisting of expression values of $N$ genes as input, where a larger survival time can be associated with a larger risk score for cases that can be ordered.

Here we have used a standard survival analysis performance assessment metric of the Concordance Index (CI) or {\it c}-index \cite{harrell1996multivariable}. It is a measure of agreement between the predicted survival risk and the observed survival. The CI is defined as follows

\begin{equation}
\text{CI} = \frac{1}{|\epsilon|} \sum_{(i,j) \in \epsilon}{1_{f(x_i)<f(x_j)}},\label{eq:01}\vspace*{-2pt}
\end{equation}
where $\epsilon = \{(i,j) \hspace{0.1cm}|\hspace{0.1cm} l_i=1 \hspace{0.1cm}\text{and}\hspace{0.1cm} t_j > t_i\}$. The CI metric is a value between $0$ and $1$, where a value of 0.5 denotes random prediction. The objective here is to maximize the CI value. Since the CI itself is not differentiable, we consider the following differentiable exponential lower bound on the CI \cite{steck2008ranking}:

\begin{equation}
\frac{1}{|\epsilon|} \sum_{(i,j) \in \epsilon}{1-e^{f(x_i)-f(x_j)}}.\label{eq:02}\vspace*{-2pt}
\end{equation}
In our regression setting, we use the negative of this lower bound as the loss function for minimization. As in \cite{wulczyn2020deep}, during optimization we ignore the denominator of Eq. \ref{eq:02}, and evaluate the loss over training batches. 

The patient cancer survival outcome endpoint used in our experiments is the disease-specific survival (DSS) endpoint \cite{liu2018integrated}. A DSS event denotes death specifically from the diagnosed cancer type, and the event time is measured from the date of initial diagnosis until that of event. The censored time denotes the time from the date of initial diagnosis until the date of death due to another cause or the date of last contact. This DSS endpoint is difficult to derive and is an approximation of the true DSS \cite{liu2018integrated}. However, we want to be able to predict the cancer-specific survival risk, which has a more direct relationship with the cancer modeling. So we have selected DSS over a more commonly used endpoint -- the overall survival (OS), and used the TCGA datasets for our experiments which provide this DSS data for multiple cancer types.

\subsection{Proposed Architecture}

\begin{figure}[!tpb]
\centerline{\includegraphics[height=67mm,width=78mm]{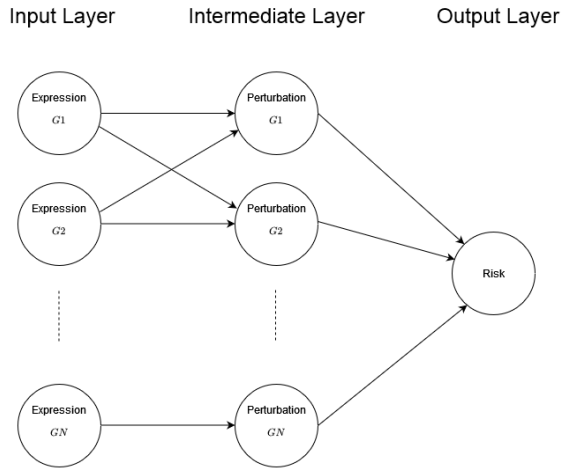}}
\caption{Proposed MPVNN architecture. Each intermediate layer neuron is assigned to represent perturbation at a gene, which is derived from its own expression and the expression of genes which are connected by either a pathway or a randomly connected edge. For example, such an edge $G1\rightarrow G2$ between two genes leads to two NN connections, one from input layer $G1$ expression neuron to intermediate layer $G2$ perturbation neuron, and second from $G2$ expression neuron to $G1$ perturbation neuron.}\label{fig:01}
\end{figure}

To solve this problem of risk prediction from gene expression profiles, we propose a simple VNN architecture with one intermediate layer connecting the input genotype layer and the output phenotype layer. The proposed MPVNN architecture is presented in Fig.~1\vphantom{\ref{fig:01}}. The input and the intermediate layers have $N$ neurons each, and the output layer has one. The connections between the input and the intermediate layers is guided by the design algorithm in Algorithm \ref{alg:MPVNN}.

\begin{algorithm}[!tpb]
	\begin{algorithmic}[1]
        \State \textbf{Input}: List of all $N$ genes $N_{all}$; List of all $P$ pathways $P_{all}$, Gene list $N_p$ \& Edge list $E_p$ for every pathway $p \in P_{all}$; Mutation data $M \in \{0,1\}^{K \times N}$;
        \State \textbf{Output}: Connection matrix $W \in \{0,1\}^{N \times N}$ denoting connections between input and intermediate layers
        \Foreach {gene $Gn \in N_{all}$}
        \State Number of mutated samples $M^{sum}_{Gn} \leftarrow \sum_{k=1}^K{M_{k,n}}$
        \EndForeach
        \State Most mutated genes $M^{mut} \leftarrow \{Gm\hspace{0.1cm}\forall\hspace{0.1cm}M^{sum}_{Gm} \geq 0.99$ quantile of all non-zero $M^{sum}_{Gn}\}$
        \Foreach {pathway $p \in P_{all}$}
        \State $frac_{p} \leftarrow \frac{|N_p|}{N}$,\hspace{0.2cm} $thr_{p} \leftarrow \frac{|N_p \cap M^{mut}|}{|M^{mut}|}$
        \EndForeach
        \State Connection matrix $W \leftarrow$ createW($N_{all}$, $N_p$, $E_p$, $thr_p$, $frac_p$ for $p \in P_{all}$)
        \item[]
        \Function{createW}{$N_{all}$, $N_p$, $E_p$, $thr_p$, $frac_p$ for $p \in P_{all}$}
        \State $W \leftarrow I_N$ 
        \State $E_{new} \leftarrow \emptyset$
        \Foreach {pathway $p \in P_{all}$}
        \Foreach {edge $a\rightarrow b \in E_p$}
        \State $rnum_1 \leftarrow random()$
        \If{$rnum_1 < thr_p$}
        \Do
        \State $rnum_2 \leftarrow random()$
        \If{$rnum_2 \geq thr_p$}
        \State $a,b \leftarrow $ Select 2 unique genes randomly from $N_p$
        \ElsIf{$rnum_2 < thr_p\frac{frac_{1\neq p}}{1-frac_p}$}
        \If{$random() \geq thr_{1\neq p}$}
        \State $a,b \leftarrow $ 2 unique genes randomly from $N_{1\neq p}$
        \Else 
        \State $a,b \leftarrow$ 2 unique genes randomly from $N_{all}-N_{1\neq p}$
        \EndIf
        \State \vdots
        \ElsIf{$rnum_2 < thr_p\frac{frac_{1\neq p} + \cdots + frac_{P\neq p}}{1-frac_p}$}
        \If{$random() \geq thr_{P\neq p}$}
        \State $a,b \leftarrow$ 2 unique genes randomly from $N_{P\neq p}$
        \Else 
        \State $a,b \leftarrow$ 2 unique genes randomly from $N_{all}-N_{P\neq p}$
        \EndIf
        \Else
        \State $a,b \leftarrow$ 2 unique genes randomly from $N_{all}-\sum_{p \in P_{all}} N_{p}$
        \EndIf
        \EndDoWhile{$(a\rightarrow b \in \cup_{p \in P_{all}} E_{p}) \lor (a\rightarrow b \in E_{new}) \lor (b\rightarrow a \in \cup_{p \in P_{all}} E_{p}) \lor (b\rightarrow a \in E_{new})$}
        \State Add $a\rightarrow b$ in $E_{new}$
        \EndIf
        \State $W_{ab} = 1$
        \State $W_{ba} = 1$
        \EndForeach
        \EndForeach
        \State \Return $W$
    \EndFunction
    \end{algorithmic}
    \caption{MPVNN design algorithm}
    \label{alg:MPVNN} 
\end{algorithm}

In the architecture, each intermediate layer neuron representing the perturbation at one gene, is connected to the input layer neuron representing the expression of the same gene. When using just pathway knowledge and no mutation data, each intermediate layer gene perturbation neuron is additionally connected to other input layer neurons representing the expression of genes which are its known neighbors in a signaling pathway. So a gene perturbation is derived from its own expression and the expression of its pathway neighbors, which are the genes with edges to or from the gene in consideration. This VNN design is unchanged by autoregulatory edges, or by cases where an edge is shared by two or more pathways or edges exist between two genes in both directions within or across pathways. This is the PVNN architecture, which is the version of the MPVNN architecture without the mutation data-based edge randomization.

The MPVNN architecture is an extension of the PVNN architecture using another type of data, non-silent gene mutation data. For a cancer type, we compute a mutation fraction per signaling pathway, which is a rough estimate of how much the signal flow in the pathway is disrupted. With the mutation fraction, we randomly replace this same fraction of the known pathway edges with new edges and connect the neurons accordingly. In this case, the intermediate layer neurons can derive perturbations at a gene from its own expression, and possibly from the expression of its neighbors as per known pathway knowledge, and possibly from the expression of other genes that it is randomly connected with, based on the mutation data, as per the design algorithm.

For the MPVNN architecture design, the input is the list of all $N$ genes and $P$ pathways, the gene and the edge lists for every individual pathway and the mutation data for the $N$ genes. We assume that over all the pathway edge lists, no autoregulatory edge exists, an edge and its reversed edge together do not exist, and an edge belongs to only one pathway. This keeps the number of connections in PVNN and MPVNN same for fair performance comparison. The design algorithm output is the matrix $W \in \{0,1\}^{N \times N}$ denoting which connections exist between the N input layer gene expression neurons and the N intermediate layer gene perturbation neurons. From the mutation data, we compute the number of samples in which each gene $Gn$ is mutated -- $M^{sum}_{Gn}$. Then we obtain $M^{mut}$ consisting of most mutated genes, each of whose above number is greater than or equal to the $0.99$ quantile of all non-zero $M^{sum}_{Gn}$. For each pathway $p$ with $frac_p$ fraction of total $N$ genes, we compute a mutation fraction $thr_p$ equal to the fraction of genes from this pathway in $M^{mut}$. Here we assume that the input gene expression profile does not consist only of genes all belonging to one single pathway, or we would use the PVNN architecture instead. For each known pathway edge, if a generated random number $rnum_1 \in [0,1)$ is $\geq  thr_p$, we select the known edge, otherwise the edge is replaced and a new edge is selected.

For replacing a known edge with a new edge, another random number $rnum_2 \in [0,1)$ is generated. If this number is $\geq thr_p$, we select two pathway $p$ genes randomly for a new edge. Otherwise, based on the value of the random number $rnum_2$, each remaining pathway $q\neq p$ is selected for consideration $thr_p\frac{frac_q}{1-frac_p}$ fraction of times. If the pathway $q$ is selected for consideration, based on whether another new random number is $\geq thr_q$, two pathway $q$ genes are selected randomly for a replacing edge, otherwise two genes are selected randomly from all genes not in pathway $q$. This is motivated by the design goal that pathways that have a high fraction of genes in most mutated genes and probably have greater disruption of signal flow, should contribute less to the genes of replacing edges. When no such pathway is selected dictated by the value of $rnum_2$, two genes, not belonging to any pathway in the list of pathways used in the design, are selected randomly for a replacing edge.

While obtaining a replacing edge in the MPVNN design, we only select two unique genes, and ensure that the selected replacing edge or its reversed edge do not belong either to the list of all known pathway edges or to the list of all already selected replacing edges. This keeps the number of connections of MPVNN same with that of PVNN, and also can help in finding important signaling connections novel to what is present in the used prior knowledge. The connection matrix $W$ is initially an identity matrix, and two entries where the row and column interchangeably denote the two genes of every selected edge, are made $1$.

Though MPVNN architecture is generalized to use prior knowledge and mutation data for multiple pathways, here as a case study, we use one well-known cancer related pathway--phosphatidylinositol 3' -kinase (PI3K)-Akt signaling pathway in its design. This pathway is a key regulator of processes involved in cell growth, proliferation, survival and apoptosis. It has been observed to play critical roles in various cancers \cite{jiang2020role}. So we have assessed the effectiveness of our proposed architecture designed with the PI3K-Akt pathway obtained from the KEGG Pathway database \cite{kanehisa2000kegg}.

\subsection{Interpretation}\label{sec:interpret}

Compared to standard black box NN survival analysis methods, a major benefit of our proposed VNN architecture is increased interpretability. Both the trained VNN architectures -- PVNN and MPVNN can be interpreted to obtain top gene sets. These sets of genes within the larger signaling pathway are linked by flow of signal that is important in the prediction of survival risk for a particular cancer type, and are ranked in order of the importance associated with a set. The benefit of MPVNN over PVNN is that from interpretation we can obtain important signaling connections which are not present in the used prior knowledge.

Here we describe a straightforward mean weight amplitude-based VNN interpretation method, which fits with the signaling edge-based design of our VNNs. We first obtain the top gene perturbations in the intermediate layer, from the absolute weights connecting the intermediate layer gene perturbation neurons to the output neuron, averaged over the VNN runs. From each top gene perturbation, we find the top gene set connected by flow of signal associated with the gene perturbations.

The first candidate gene is the one with the top gene perturbation, and then the following process is repeated for every new gene added to the top gene set. For every candidate gene, we evaluate which of the gene expression neurons in the input layer, apart from the candidate gene itself and the last gene added to the top gene set in consideration, has the highest mean absolute weight connecting to the candidate gene perturbation neuron in the intermediate layer. We check whether the highest weight is $\geq 0.85$ quantile of all the non-zero weights between the input layer and the intermediate layer neurons. If this new gene already belongs to the top gene set in consideration, we stop the process. Otherwise, we check whether the weight connecting the expression neuron for the newly selected gene to its perturbation neuron in the intermediate layer satisfies the weight threshold above. We additionally check if the perturbation neuron for the newly selected gene to the output layer phenotype neuron is $\geq 0.85$ quantile of all the non-zero weights between the intermediate layer and the output layer neurons. Then this new gene is added to the top gene set. This whole process is done twice starting from the top gene perturbation to be able to possibly capture the important signal flow into and out of the gene.

The thresholds used are applied to ensure that genes with low connection weights and hence lower importance are not included in the top gene sets. We used a value of $0.85$ quantile, which for the intermediate layer neurons, roughly translates to that out of the 1440 gene perturbations, around 200 can play some role of importance in risk prediction for a cancer type. The top gene sets obtained from the interpretation would depend on the values of the thresholds used.


\section{Experiments}\label{sec:datasets}

Our experimental TCGA data for 10 cancer types is obtained from the UCSC Xena browser \cite{goldman2020visualizing}. The data includes gene expression RNAseq data, binary gene-level non-silent mutation data and disease-specific survival data. We consider cancer samples, one per patient, which do not fall under the normal category in TCGA. For the expression profile, apart from the genes in the PI3K-Akt pathway, we also consider 1285 genes found to have systematic expression change in cancer \cite{torrente2016identification}. Finally our patient expression profile consists of 1440 genes, for which both expression and mutation data are available.

The cancer types used in our experiments are presented in detail in Table \ref{datasets:tbl1}. For each cancer type, the input-output data is split after a random shuffle into training ($80\%$) and test ($20\%$) sets with stratification on the censoring indicator/ event observed values. For the MPVNN architecture design, the available mutation data is used as a whole. For every machine learning method in our experiments, the input data features are standardized by mean subtraction and scaling to unit variance, based on those metrics computed from the training data.

\begin{table*}[!t]
\caption{Our experimental datasets.}
\begin{tabular}{| p{18mm} |p{73mm} |p{15mm} |p{24mm}|p{13mm}|}\toprule
Cancer & Description & Number of samples & Events observed & Censored \\ \midrule
BLCA & Bladder urothelial carcinoma & 390 & 119 & 271 \\
BRCA & Breast invasive carcinoma & 1070 & 83 & 987\\
COADREAD & Colon and rectum adenocarcinoma & 354 & 41 & 313 \\
GBM & Glioblastoma multiforme & 147 & 113 & 34 \\
HNSC & Head and neck squamous cell carcinoma & 493 & 130 & 363\\
KIRC & Kidney renal clear cell carcinoma & 522 & 109 & 413 \\
LIHC & Liver hepatocellular carcinoma & 361 & 79 & 282 \\
LUNG & Lung squamous cell carcinoma and adenocarcinoma & 914 & 199 & 715 \\
OV & Ovarian serous cystadenocarcinoma & 272 & 150 & 122 \\
STAD & Stomach adenocarcinoma & 386 & 95 & 291 \\\bottomrule
\end{tabular}
\label{datasets:tbl1}
\end{table*}

We evaluated the performance of MPVNN against other comparable NN architectures. First there is the PVNN architecture which is designed using pathway knowledge and no mutation data-based randomization. In RaNN which is another randomized version of PVNN, the connections between the hidden layer and the input layer neurons are randomly shuffled. This represents a same sized NN with the same number of connections but designed without using prior knowledge or additional mutation data. We evaluated the performance of the fully connected artificial neural network or ANN with the same number of neurons. We have also compared the performance of these NN methods with a standard survival analysis method -- the semi-parametric Cox Proportional Hazards (Cox-PH) model \cite{cox1972regression}. We drop the expression values of those genes which give an initial warning of having very low variance in the Cox PH regression fitter.

We used a 4-fold cross-validation on the training set to identify the optimal hyperparameters. The optimal hyperparameter setting for a particular method for a particular cancer type is selected from a certain number of random searches on the list of given hyperparameter values. For this number, we used around $10\%$ of the total number of possible combinations. So the number of random searches was set to be 300 for NNs and 10 for Cox-PH model. The list of values to choose from for NNs is given as follows :

\begin{enumerate}
    \item Activation function : tanh, sigmoid, ReLU
    \item Optimizer : Adam, SGD
    \item Learning Rate : 0.1, 0.01, 0.001
    \item Batch Size : 16, 32, 64, 128
    \item Epoch : 10, 25, 50, 100
    \item Regularizer : L1, L2
    \item Regularizer parameter : 0.0, 1e-5, 1e-3, 1e-2, 1e-1, 1.0.
\end{enumerate}
The hyperparameters for the Cox-PH model to select from are penalizer values in the set $$\{0.0,0.0001,0.0005,0.001,0.005,0.01,0.05,0.1,0.5,1.0\}$$ and the L1 ratio values between $0$ and $1$ in intervals of $0.1$. For randomized architectures MPVNN and RaNN, the optimal connection matrix is also selected from the same 300 random searches.

The performance metric used is the CI value on the hold-out test set. We present the mean and the standard deviation from 20 runs. The data and code are available at \url{https://github.com/gourabghoshroy/MPVNN}.

\section{Results}

In Table~\ref{Tab:01} we present the results of our MPVNN architecture for cancer-specific survival risk prediction, for each cancer type, and finally these values macro-averaged over all cancers. We also show each method's macro-average weighted rank, where the weighted rank is calculated as the difference between the maximum of mean CI metric values for all methods and the mean CI metric value for a particular method in a cancer type. The lower the macro-average weighted rank, the closer to the best mean performance a method's mean performance is on an average across cancer types.

\begin{table*}[!t]
\caption{Cancer-specific survival risk prediction performance evaluation of MPVNN. The mean and standard deviation of the CI metric from 20 runs are shown for each cancer type, and then these values are macro-averaged over all cancer types. The macro-average weighted ranks are also presented. The mean best, that is the method with the highest mean value, individually and macro-averaged, and the one with the lowest macro-average weighted rank, are all marked in boldface.} 
\begin{tabular}{| p{30mm} |p{22mm} |p{22mm} |p{22mm}|p{22mm}|p{22mm}|}\toprule
Cancer & Cox-PH & ANN & RaNN & PVNN & MPVNN\\\midrule
BLCA & 0.6919$\pm$0.0000 & 0.7455$\pm$0.0107 & \textbf{0.7463$\pm$0.0046} & 0.7130$\pm$0.0089 & 0.7214$\pm$0.0029 \\
BRCA & 0.4810$\pm$0.0142 & \textbf{0.6507$\pm$0.0174} & 0.5434$\pm$0.0082 & 0.6367$\pm$0.0151 & 0.6298$\pm$0.0040 \\
COADREAD & 0.5868$\pm$0.0000 & 0.7731$\pm$0.0673 & 0.7658$\pm$0.0474 & 0.6085$\pm$0.0230 & \textbf{0.7804$\pm$0.0267} \\
GBM & \textbf{0.5822$\pm$0.0041} & 0.5153$\pm$0.0521 & 0.5762$\pm$0.0421 & 0.4991$\pm$0.0168 & 0.5763$\pm$0.0115 \\
HNSC & 0.4912$\pm$0.0006 & 0.5864$\pm$0.0092 & 0.5920$\pm$0.0091 & \textbf{0.6498$\pm$0.0025} & 0.5827$\pm$0.0081 \\
KIRC & 0.7747$\pm$0.1159 & 0.8374$\pm$0.0257 & \textbf{0.8474$\pm$0.0063} & 0.7955$\pm$0.0049 & 0.8082$\pm$0.0063 \\
LIHC & 0.5549$\pm$0.0042 & 0.7527$\pm$0.0292 & 0.7600$\pm$0.0060 & \textbf{0.7918$\pm$0.0052} & 0.7878$\pm$0.0050 \\
LUNG & \textbf{0.6644$\pm$0.0000} & 0.5837$\pm$0.0124 & 0.5649$\pm$0.0035 & 0.6248$\pm$0.0031 & 0.6160$\pm$0.0057 \\
OV & 0.5895$\pm$0.0305 & 0.5886$\pm$0.0334 & 0.5869$\pm$0.0130 & 0.6102$\pm$0.0068 & \textbf{0.6248$\pm$0.0050} \\
STAD & \textbf{0.7429$\pm$0.0000} & 0.6207$\pm$0.0516 & 0.6699$\pm$0.0087 & 0.5502$\pm$0.0395 & 0.6800$\pm$0.0062 \\ \hline
MACRO-AVERAGE & 0.6160$\pm$0.0170 & 0.6654$\pm$0.0309 & 0.6653$\pm$0.0149 & 0.6480$\pm$0.0126 & \textbf{0.6807$\pm$0.0081} \\
MACRO-AVERAGE WEIGHTED RANK & 0.0921 & 0.0427 & 0.0428 & 0.0601 & \textbf{0.0273} \\\bottomrule
\end{tabular}
\label{Tab:01}
\end{table*}

Our results show that the overall mean cancer-specific risk prediction performance of MPVNN is the best among the methods that were tested. Firstly, it is much much better than the standard Cox-PH survival analysis model macro-averaged across all cancer types. Second, its overall performance is better than those of other comparable NN architectures -- PVNN, RaNN and ANN. Compared to the next best method, MPVNN on macro-average has a mean CI metric higher by $0.015$. Even though MPVNN is not the best in all of the 10 cancer types, its macro-averaged metrics demonstrate that the incorporation of signaling pathway knowledge and gene mutation data-based edge randomization in the VNN design can improve the overall prediction performance.

\begin{figure*}[!htpb]
\centerline{\includegraphics[height=103mm,width=133mm]{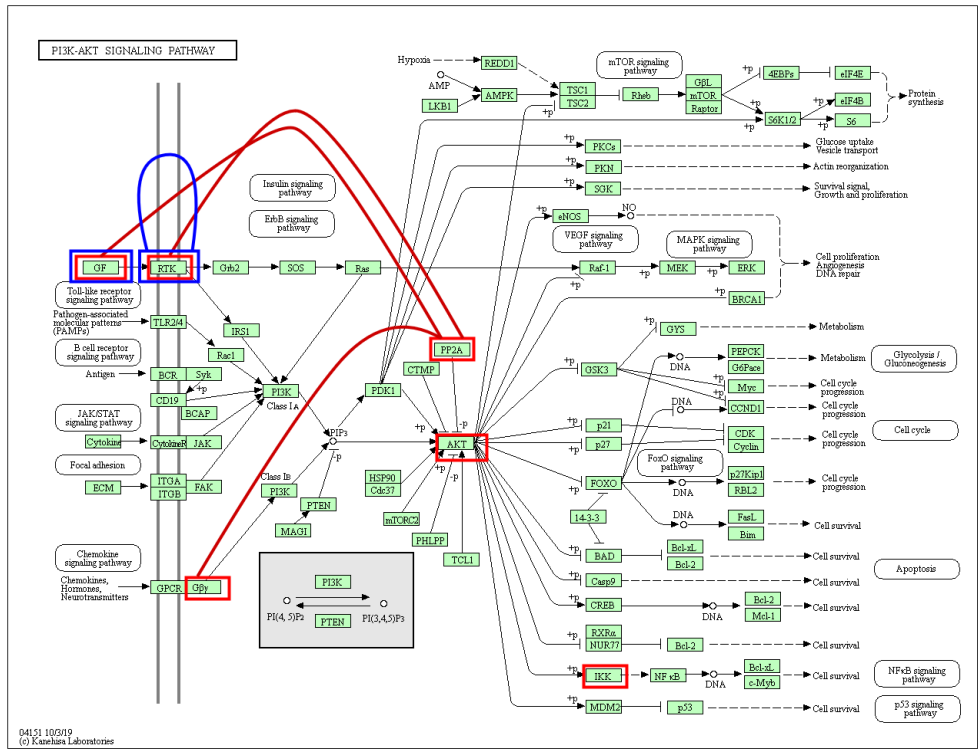}}
\caption{Top gene set from MPVNN interpretation for Ovarian (OV) and Liver (LIHC) cancers. Ovarian : GNB3--PPP2R1B--FGF2$\rightarrow$\textbf{NGFR}--PPP2R2B$\rightarrow$AKT1$\rightarrow$CHUK, Liver : ANGPT2$\rightarrow$FLT3--\textbf{FLT4}$\leftarrow$ANGPT2. The gene group in the PI3K-Akt pathway diagram \cite{kanehisa2000kegg} for each gene in the top gene set is bordered by a colored rectangle (Ovarian - Red, Liver - Blue). The novel connections which are not present in the pathway knowledge are shown by colored curved lines.}\label{fig:02}
\end{figure*}

To demonstrate the interpretability of our proposed MPVNN architecture, we have shown two top gene sets obtained from MPVNN interpretation in Fig.~2\vphantom{\ref{fig:02}}. The top gene sets are given for 2 cancer types, ovarian and liver, in which the VNNs outperform other methods (MPVNN performs the best and the second best). We have selected the top gene set which has the first occurrence of any signaling connection that is not present in the used PI3K-Akt pathway edge list. For ovarian cancer, the top gene set consists of these genes : GNB3--PPP2R1B--FGF2$\rightarrow$\textbf{NGFR}--PPP2R2B$\rightarrow$AKT1$\rightarrow$CHUK, where the top gene perturbation is marked in bold and the novel signaling interactions are marked by dashes. The shown top gene set for liver cancer comprises ANGPT2$\rightarrow$FLT3--\textbf{FLT4}$\leftarrow$ANGPT2. An important novel connection between genes FLT4 and FLT3 in the same gene group RTK is interpreted from the MPVNN architecture, however these 2 genes are not connected by any given PI3K-Akt signaling path.

\section{Discussion}

From our results, we demonstrate that the overall cancer-specific survival risk prediction mean performance of our proposed MPVNN architecture is substantially better than other survival analysis methods. However, the mean performance of MPVNN is not the best in every single cancer type, given the corresponding data. The Cox-PH model is found to be a better risk prediction model than the NNs in some cancer types, though MPVNN is observed to be much better on macro-averaged metrics. The high importance of the PI3K-Akt pathway in risk prediction is highlighted for those particular cancer types where PVNN performs the best. We also observe that for some cancers, the randomized RaNN or the fully connected ANN gives the best mean CI metrics. Since these two architectures lack interpretability, it is more difficult to understand in biological terms how they work. When RaNN is the best, one possibility is that there are other pathways involving the input genes, which are more effective in risk prediction. Incorporating knowledge and mutation data for multiple pathways in the architecture design might be helpful for performance. As described previously, our VNN architectures can be applied to use multiple pathways. For the cancer where ANN is the best, one possibility is that pathways which can play significant roles in risk prediction are not well represented in the input dataset genes, and a larger set of input genes, along with multiple pathway knowledge and mutation data in the architecture design, might improve the performance. 

A major benefit of our MPVNN architecture is increased interpretability compared to black box NN methods, making it more reliable for use in clinical survival analysis. To validate some of the insights we obtained from MPVNN interpretation, we looked for literature evidence. In the top gene set for ovarian cancer, we observed important signaling interactions within the PI3K-Akt pathway between parts of the growth factor, the chemokine, the Akt and the NF$\kappa$B signaling pathways. Some of these interactions are novel to what exists in the used prior knowledge. Signal flow involving these four pathways together being important in the context of higher mortality in ovarian cancer has been previously indicated \cite{dong2013cxcr2,son2013characteristics}.

In the top gene set in liver cancer, we observed important pathway edges ANGPT2$\rightarrow$FLT3 and ANGPT2$\rightarrow$FLT4 and a novel interaction between RTK genes FLT3 and FLT4. Connection to ANGPT2 from a RTK gene via calcineurin-NFAT has been studied in lung metastasis \cite{minami2013calcineurin}. So such a signaling path from any of the RTK genes FLT3 or FLT4 back to ANGPT2 can correspond to the interpreted novel interaction. Both the FLT3 and FLT4 genes are among the targets of the popular drug Sorafenib, which has been used effectively in the treatment of advanced liver cancer for over a decade \cite{marisi2018ten}. Interestingly, in an extensive Sorafenib trial \cite{llovet2012plasma}, ANGPT2 was found to be an independent survival predictor in the overall and the drug treated population, but not related with drug induced survival benefit. Compared to the baseline concentration, ANGPT2 concentration after 12 weeks of treatment was observed to increase in the placebo group, however it remained almost unchanged for treatment with the drug. So it is possible that the signal flow in the MPVNN interpreted top gene set is disrupted during treatment with Sorafenib by targeting both FLT3 and FLT4 genes, as a result of which the ANGPT2 concentration remains almost constant. Later the signal flow involving these genes probably again comes into effect.

Based on the above literature validation, we argue that MPVNN interpreted insights are reliable, pointing to signal flow that have critical roles in controlling cancer-specific survival risk. Further validation of these insights would require experimental verification, which is outside the scope of this study. These insights are flexible enough to include important signaling connections that are not present in the used prior knowledge. MPVNN interpreted insights can have correspondence with the mode of action of existing drugs, and can provide directions for novel and more effective single or combinatorial drug therapy.

Edge randomization based on additional mutation data, which models how a pathway structure can change for a particular cancer type and replaces known edges with new connections, is a novelty of our MPVNN architecture. Heteroscedastic dropout with privileged or additional information is used in \cite{lambert2018deep}, although this differs substantially from MPVNN. Dropout for a NN architecture is used only during learning and not in prediction, whereas the edges in the MPVNN architecture are randomized at the beginning, before learning and subsequent prediction. Also in MPVNN, the use of the additional mutation data in the supervised prediction setting is unsupervised, without the requirement of being related to the input-output training data. This makes the architecture more robust to bias that may arise in supervised machine learning when some population groups are not well represented in the training set, for instance due to lack of survival data here, but mutation data is available for those groups and can be incorporated in the model.

As only the PI3K-Akt pathway used in this study is not sufficient in fully demonstrating MPVNN predictive power, future work would look at conducting larger scale experiments with larger gene expression profiles using knowledge and data for multiple pathways in the MPVNN architecture design to further evaluate cancer-specific survival risk prediction performance. In future, we would also like to experimentally investigate the roles of the important signal flow identified from MPVNN interpretation for different cancer types, and importantly explore their relevance to cancer drug target identification.

\section*{Funding}
Gourab Ghosh Roy is supported by a Priestley Scholarship for joint study at the University of Birmingham and the University of Melbourne.

\bibliographystyle{unsrt}  
\bibliography{references}

\end{document}